\documentclass[12pt]{article}
\usepackage{cite}

\def\beq{\begin{equation}}
\def\eeq{\end{equation}}
\def\bea{\begin{eqnarray}}
\def\eea{\end{eqnarray}}
\newcommand{\gsim}{\lower.7ex\hbox{$\;\stackrel{\textstyle>}{\sim}\;$}}
\newcommand{\lsim}{\lower.7ex\hbox{$\;\stackrel{\textstyle<}{\sim}\;$}}

\begin{document}

\thispagestyle{empty}
\vspace*{.5cm}
\noindent
DESY 02-218 \hspace*{\fill} December 12, 2002\\
\vspace*{1.6cm}

\begin{center}
{\Large\bf Power-Like Threshold Corrections to\\[.3cm] Gauge Unification
in Extra Dimensions}
\\[2.0cm]
{\large A. Hebecker and A. Westphal}\\[.5cm]
{\it Deutsches Elektronen-Synchrotron, Notkestrasse 85, D-22603 Hamburg,
Germany}
\\[1cm]

{\bf Abstract}\end{center}
\noindent
One of the much-debated novel features of theories with extra dimensions is 
the presence of power-like loop corrections to gauge coupling 
unification, which have the potential of allowing a significant reduction of 
the unification scale. A recognized problem of such scenarios is the UV
sensitivity of the above power corrections. We consider situations where 
the grand unified group is broken by the vacuum expectation value of a bulk
field and find that, because of the softness of this extra-dimensional
symmetry breaking mechanism, power-like threshold corrections are 
calculable and generic in many of the most relevant settings. While the 
precision is limited by the presence of higher-dimension bulk operators, 
the most dangerous of these operators are naturally forbidden by 
symmetries of the bulk theory. Particularly interesting and constrained 
scenarios arise in the context of higher-dimensional supersymmetry. Our 
phenomenological exploration of SU(5) models in 5d, linked in particular 
with more recently discussed orbifold GUT models, shows promising results.

\newpage

\section{Introduction}
Grand unified theories provide an elegant explanation of the fermion quantum
numbers of the Standard Model (SM)~\cite{gg} (also~\cite{ps}). Together
with the success of gauge coupling unification~\cite{gqw} in supersymmetric
(SUSY) extensions of the SM~\cite{drw}, this has established high-scale 
grand unification as the standard framework for the discussion of physics
above the electroweak scale.

During the last few years, the above paradigm has been challenged by various
scenarios with extra dimensions compactified at scales below $M_{\rm
GUT}\sim 10^{16}$ GeV. In particular, Dienes, Dudas and 
Gherghetta~\cite{ddg} have proposed low-scale gauge unification as a 
possible consequence of power-like loop corrections to gauge coupling 
constants~\cite{tv,rob}. However, objections to this proposal
have been raised on the basis that the relevant loop corrections are
completely UV dominated and that, as a result, no precise statement about
the ratio of low-energy gauge couplings can be made without a UV completion
of the higher-dimensional SM-like theory (see, e.g.,~\cite{obj}). The issue 
of `power law running' was also discussed in connection with 
`deconstruction' and warped 5d models (see, e.g.,~\cite{dec} 
and~\cite{gr,ads}). In the present paper we demonstrate that, if the 
GUT group is softly
broken in the weak-coupling regime of the higher-dimensional theory, the
resulting power-like threshold corrections can be numerically important,
calculable, and of universal nature.

To be specific, let us first consider $d$-dimensional pure Yang-Mills
theory, compactified to 4 dimensions on a torus of radius $R$. Scattering
processes in the 4d theory at energies near the compactification scale
$M_c\sim 1/R$ can be used to define a 4d gauge coupling $\alpha_4(M_c)
=g_4^2(M_c)/(4\pi)$. In the following, this quantity will be considered as
the basic physical observable of the low energy effective theory. It is
linked to processes at energies far below $M_c$ by conventional 4d
logarithmic running. The relation to the coupling constant $\alpha_d$ of
the $d$-dimensional theory is given by
\beq
\alpha_4(M_c)^{-1} \sim \alpha_d(\mu)^{-1}R^{d-4}+
f_{\mbox{\scriptsize 1-loop}}(\mu,R)+\mbox{higher orders}\,,\label{tree}
\eeq
where $\mu$ characterizes the renormalization point of the
higher-dimensional field theory (see, e.g.,~\cite{wei}). For $\mu\gg M_c$,
the leading contribution from $f_{\mbox{\scriptsize 1-loop}}$ is $\sim
(\mu R)^{d-4}$. It describes the power-divergent loop-correction to the
$F^2$ term in the bulk, multiplied by the extra-dimensional volume. Since
the l.h. side is $\mu$-independent, we have $\alpha_d(\mu)^{-1}\sim
M^{d-4}-\mu^{d-4}$, where $M$ can be considered as the fundamental UV
scale of the model, and ${\cal O}(1)$ numerical coefficients (which
depend on the renormalization scheme) have been suppressed. It is
convenient to assume $\mu\ll M$, so that $\alpha_d\sim M^{4-d}$.

Next, we assume that the vacuum expectation value (VEV) of a bulk Higgs
breaks the simple gauge group $G$ of the fundamental theory to a subgroup 
$H=H_1\times\cdots\times H_n$ (which is a direct product of simple groups 
and U(1) factors). The Higgs breaking is characterized by an energy scale 
$M_B$, related to the masses of vector bosons and physical scalars. For 
$M_c\ll M_B\ll\mu\ll M$, the 4d gauge couplings, labelled by the index 
$i=1...n$, are now given by
\bea
\alpha_{4,i}(M_c)^{-1} &\sim & \alpha_d(\mu)^{-1}R^{d-4}+(\mu R)^{d-4}+
f_{\mbox{\scriptsize 1-loop},i}(\mu,R,M_B)+\mbox{higher orders}\,.
\label{ali}
\eea
Here we have split the 1-loop correction into a universal ($i$-independent)
piece carrying the leading divergence $\sim \mu^{d-4}$ and the non-universal 
piece $f_{\mbox{\scriptsize 1-loop},i}$. 
To understand this structure, it is sufficient to observe that, while
the bulk theory at energies below $M_B$ possesses non-universal (with
respect to $i$) power divergences of degree $d-4$, such divergences can not 
be present in the unbroken high-scale theory. Thus, their contribution
to the coefficients of $F_i^2$ is suppressed by $M_B^2$. To be more 
specific, the function $f_{\mbox{\scriptsize 1-loop},i}$ may be considered 
as arising from differences of one-loop integrals with massive and 
massless vector bosons,
\beq
\int^\mu\frac{d^dk}{(k^2+M_B^2)^2}-\int^\mu\frac{d^dk}{(k^2)^2}\sim
M_B^2\mu^{d-6}\,,
\eeq
which demonstrates the structure of the $M_B$-suppression. This 
estimate is, however, only valid for $d>6$. For $d=5$ this term is finite 
and calculable, so that Eq.~(\ref{ali}) has to be replaced by
\beq
\alpha_{4,i}(M_c)^{-1} \sim \alpha_5(\mu)^{-1}R+\mu R+c_iM_BR+\cdots\,.
\label{ali5}
\eeq
This structure was previously discussed in U(1) toy models~\cite{gr,ads}. 
Except for the non-universal numbers $c_i$, numerical coefficients 
have been suppressed in the above estimates. Furthermore, both higher-loop 
and volume-suppressed terms have been dropped in Eqs.~(\ref{ali}) 
and~(\ref{ali5}). 

For $d=6$, the $M_B$ suppressed term reads $c_i(M_BR)^2\ln(\mu/M_B)$. 
This means that non-universal counterterms (corresponding to 
higher-dimension operators) have to be present for consistency of the 
theory. Thus, although an ${\cal O}(1)$ term coming with the log remains 
undetermined, the coefficients $c_i$ and therefore the log-enhanced piece
is calculable. 

The above contributions proportional to $c_i$ provide corrections to 
$\alpha_{d,i}^{-1}$ of relative size $(M_B/M)^{d-4}$. At first sight, the 
phenomenological relevance of these corrections appears to be limited by 
possible higher-dimensional operators, e.g., tr$[F^2\cdot\Phi]$ (where 
$\Phi$ is the bulk field developing a symmetry-breaking VEV). In principle, 
such operators can give rise to non-universal corrections as large as the 
loop-effects discussed above.\footnote{
This has been pointed out in~\cite{gr} in the context of a 5d toy model GUT 
with gauge group U(1)$\times$U(1)$'$.}
However, as we will explain in detail below, in the simplest and most popular
higher-dimensional scenarios, the leading dangerous operators are either
automatically forbidden or can be forbidden by minimal symmetry assumptions.
Furthermore, it turns out that the coefficients $c_i$ are governed by the
basic group theoretic structure of the theory and are therefore fairly
model-independent. Thus, we conclude that power-like threshold corrections
to gauge unification can and should be taken seriously at a quantitative
level.

In Sect.~\ref{cal}, we consider higher-dimensional Yang-Mills theory with a
bulk Higgs field $\Phi$. Given an appropriate bulk potential, $\Phi$ will
develop a symmetry breaking VEV leading to vector bosons with masses 
$\sim M_V$ and a number of physical scalars with masses $\sim M_S$ (the 
scalar mass spectrum 
depends on the parameters of the potential). For $d=5$, it suffices to
forbid operators linear in $\Phi$ by a $Z_2$-symmetry to obtain potentially
sizeable, controlled power-corrections. Furthermore, in sufficiently flat
potentials the vector mass will dominate the scalar masses, so that $M_S\ll
M_B$ and the corrections are independent of the GUT-Higgs potential.

In Sect.~\ref{sup}, we discuss the supersymmetric theory. Higher-dimensional 
supersymmetry strongly constrains the possibilities of gauge symmetry 
breaking by a Higgs field and the arising loop corrections. In 
particular, breaking by an adjoint Higgs hypermultiplet does not lead to 
power-corrections because the structure of the model is that of $N=4$ 
super-Yang-Mills theory. However, potentially large and calculable 
corrections arise from breaking by the adjoint scalar of the 5d vector 
multiplet.

In Sect.~\ref{su5}, realistic 5d SU(5) versions of the above generic GUT
scenario are discussed. Lowering the compactification scale significantly
below $10^{16}$ GeV potentially leads to fast proton decay which can,
however, be avoided by working on an orbifold ($S^1/Z_2$ or $S^1/(Z_2\times
Z_2')$), where SU(5) is not a good symmetry on at least one of the branes.
(This idea, already discussed in the present context in~\cite{ddg}, has
recently been extensively used in the context of orbifold
GUTs~\cite{kaw,af,hn,hmr}. For larger gauge groups and more than
5 dimensions see, e.g.,~\cite{d6}.)
Given the explicit SU(5) breaking on the brane, it is quite natural that
an adjoint bulk scalar develops a VEV in U(1)$_Y$ direction. The relative
size of the resulting power-like contributions to the U(1), SU(2), and
SU(3) couplings is governed by the Casimirs of the respective adjoint
representations. Thus, their effect mimics the dominant (pure gauge) part
of conventional logarithmic running. Additional bulk fields can, if they
acquire non-universal masses because of the symmetry breaking adjoint
bulk VEV, contribute further, model-dependent, power law corrections.

Our conclusions are given in Sect.~\ref{con}

\section{Calculable bulk threshold corrections}\label{cal}

Let us begin by considering a $d$-dimensional Yang-Mills theory with simple 
gauge group $G$ and a Higgs-field $\Phi$ transforming in some representation 
of $G$. The lagrangian reads
\beq
{\cal L}=-\frac{1}{2 g_d^2}\cdot \mbox{tr}\left(F_{MN}F^{MN}\right)-
\left(D_M\Phi\right)^\dagger\left(D^M\Phi\right)-V(\Phi)\,,\label{Leff}
\eeq
where $F_{MN}$ is the field strength tensor, $D_M$ is the covariant 
derivative, and the indices $M,N$ run over 0,...,3,5,...,$d$. We assume 
that $\Phi$ develops a VEV breaking $G$ to a subgroup $H=H_1\times\cdots 
\times H_n$. Without supersymmetry, this can simply be realized by choosing 
an appropriate bulk potential $V(\Phi)$. Higher-dimensional supersymmetry 
restricts possible bulk interactions and different origins for a bulk VEV 
have to be considered (cf.~Sects.~\ref{sup} and~\ref{su5}).

At tree level, the couplings $\alpha_{d,i}$ 
of the group factors $H_i$ are equal to the coupling $\alpha_d$ of $G$. 
At one loop, one has to calculate the contributions of the light and heavy 
vector bosons and the physical Higgs scalars to the coefficients of the 
$F_i^2$ terms, i.e., to the normalization of the field-strength terms of 
the unbroken subgroup factors. This calculation was done in the context of 
4d GUTs in dimensional regularization~\cite{wei1,thresh} (see 
also~\cite{fra}), 
so that the $d$-dimensional result can simply be taken from~\cite{wei1}:
\beq
\alpha_{d,i}^{-1}=\alpha_d^{-1}+\frac{\Gamma(2-d/2)}{6(4\pi)^{d/2-1}}
\left[-(25-d)\sum_{r_i} M_{V,{r_i}}^{d-4}T_{r_i}+\sum_{r_i'} 
M_{S,r_i'}^{d-4}T_{r_i'}+2s_d\sum_{r_i''} M_{F,r_i''}^{d-4}T_{r_i''}
\right]\,.\label{wf}
\eeq
Here $r_i,r_i'$ and $r_i''$ label the representations under $H_i$ of the 
vector, scalar and spinor particles and $M_{V,{r_i}}$, $M_{S,r_i'}$, and 
$M_{F,r_i''}$ stand for the corresponding masses. (Although the minimal 
setting discussed at the moment has no fermions, we have included a possible 
fermionic contribution into the above equation for completeness. The number 
$s_d$ characterizes the dimension of the relevant spinor.) Furthermore, 
$T_{r_i}$ is defined by tr$[T^aT^b]=\delta^{ab}T_{r_i}$, where $T^{a,b}$ are 
the generators in the representation $r_i$ (and analogously for $r_i'$ and 
$r_i''$). 

Concerning the structure of Eq.~(\ref{wf}), several comments are in order. 
We have chosen $\alpha^{-1}$ (rather than $\alpha$ or $g$) as our basic
quantity because it can be interpreted as the coefficient of the $F^2$
operator and hence the further transition to the 4d theory proceeds simply
by multiplication with the volume factor. Of course, this direct relation 
between Eq.~(\ref{wf}) and Eq.~(\ref{ali}) works only up to terms 
suppressed by a volume factor. We will discuss such terms in more detail 
below. Note furthermore that Eq.~(\ref{wf}) does not contain contributions 
from the gauge bosons of the unbroken subgroup. In our context, the reason 
for this is the masslessness of these vector bosons. Because of the 
absence of a mass scale, the corresponding loop integrals have a pure 
power of the loop momentum in the integrand and therefore vanish in 
dimensional regularization.\footnote{
This argument works only for $d>4$, where the coupling can be defined at
zero external momentum. In 4 dimensions, the relevant loop integrals 
require the external momentum as IR regulator, and as a result the familiar 
contribution $\sim\ln(\mu^2/Q^2)$ from massless gauge bosons appears.}

By power counting, we expect the one-loop correction to $\alpha_{d,i}^{-1}$ 
to diverge with the $(d-4)$th power of the cutoff. The fact that this does 
not show up in Eq.~(\ref{wf}) is due to the use of dimensional 
regularization. However, this does does not restrict the validity of our 
conclusions in any way. On the one hand, this leading power-divergence is 
$G$-universal (independent of $i$) because of the symmetric structure of 
the UV theory and can thus be absorbed in a redefinition of $\alpha_d^{-1}$. 
On the other hand, the main phenomenological implications depend only on 
the differences between the inverse gauge couplings of the group factors 
$H_i$ and are therefore not affected by a $G$-universal correction. 

The further analysis depends crucially on two closely related issues: the 
possible existence of higher-dimension operators that can compete with the 
corrections on the r.h. side of Eq.~(\ref{wf}) and the UV divergences that 
are present even in these non-$G$-universal loop corrections. To be 
specific, the lagrangian generically contains terms
\beq
\sim \frac{1}{M^k}\Phi^n (F_{MN})^2\,,\label{hdo}
\eeq
where we have assumed that the relevant product of representations of $G$ 
contains a singlet, and $k$ is chosen to ensure the overall mass dimension 
$d$. When $\Phi$ develops a VEV $v=M_V/g_d$ (vector boson masses are 
generated as in the familiar 4-dimensional setting), the operator in
Eq.~(\ref{hdo}) can lead to non-universal corrections to 
$\alpha_{d,i}^{-1}$ at tree-level. Since $g_d^2\sim M^{4-d}$ (the 
fundamental UV scale of the theory), the relative size of this correction
is given by 
\beq
\frac{\Delta\alpha_{d,i}^{-1}}{\alpha_d^{-1}}\sim \left(\frac{M_V}{M}
\right)^n\,.\label{del}
\eeq
This has to be compared to the correction from Eq.~(\ref{wf}) which, 
focussing on the vector boson part, is of relative size $(M_V/M)^{d-4}$. 
Given the possibility that $n=1$, this appears to be discouraging. 
However, it is important to keep in mind that certain values of $n$ may 
be forbidden by group theory or other symmetries. For example, for $d=5$ 
the leading calculable correction is of relative size $M_V/M$ $(M_V/M\ll 
1)$, and a simple $Z_2$ symmetry $\Phi\to -\Phi$ is sufficient to forbid 
the competing $n=1$ term from Eq.~(\ref{del}). The $n=2$ term is of 
relative size $(M_V/M)^2$ and therefore negligible. 

Higher-dimension operators can act as counter terms and are therefore 
intimately linked to the divergence structure of the non-universal 
corrections on the r.h. side of Eq.~({\ref{wf}). In 5d, the non-universal 
part of the loop correction is finite, which is consistent with the 
possibility of forbidding the relevant operator by a $Z_2$ symmetry. 
For $d\to 6$, the Gamma function develops a pole, showing that the 
non-universal term is afflicted by a logarithmic divergence. Although this 
implies the existence of a higher-dimension operator $\sim F^2\Phi^2$
providing the counter term, predictivity is maintained at the leading 
logarithmic level. To be specific, we assume that the 
divergence is cured by a theory of higher symmetry at the scale $M$ and 
that there are no anomalously large non-universal threshold effects 
associated with this transition. In short, we work at leading-log 
approximation in $M/M_V$. This logarithm can be extracted from 
Eq.~(\ref{asu}), as is common in 4d, by setting $d=6-2\epsilon$, introducing 
appropriate factors $\mu^{2\epsilon}$ to keep the correct dimensionality, 
expanding in $\epsilon$, and letting $\epsilon\to 0$ and $\mu\to M$. 
Focussing on the vector contribution, the result reads
\beq
\alpha_{4,i}^{-1}(M_c)=V\alpha_6^{-1}+\frac{1}{3(4\pi)^2}\,
19\sum_{r_i} (VM_{V,r_i}^2)T_{r_i}\ln\frac{M}{M_{V,r_i}}\,.
\eeq
This should provide a good description if $M_c^2\ll M_V^2\ll M^2$ and 
scalar and fermion masses are small. 

In more than 6 dimensions, power-counting suggests that there are 
non-universal power-divergences. More specifically, the explicitly 
calculated logarithmic divergence in $d=6$ suggests a power-divergence of 
degree $d-6$ in $d$ dimensions, which would have to come with a factor 
$M_V^2$ for dimensional reasons. The corresponding counter term is
provided by the operator $\Phi^2F^2$, which should therefore always be 
included in the lagrangian. The term $\sim M_V^{d-4}$ in Eq.~(\ref{wf}) is
subdominant with respect to this operator. Thus, quantitative statements
depend on a more detailed knowledge of the UV structure of the 
theory. However, it is likely that the group-theoretical 
specification of the VEV of $\Phi$ and a classification of the singlets 
contained in $\Phi^2F^2$ will be sufficient to uniquely determine or
strongly constrain the way in which $\Phi^2F^2$ terms contribute to 
gauge coupling differences $\alpha_{4,i}^{-1}(M_c)-\alpha_{4,j}^{-1}(M_c)$. 

Dangerous higher-dimension operators mixing $\Phi$ and $F^2$ may also 
reside on branes. But in this case their contribution to the observed 
effective 4d couplings is further suppressed by volume factors. For 
example, the contributions of operators on branes of co-dimension $d_c$ are
suppressed by the potentially small factor $(MR)^{-d_c}$, where $R$ is 
the compactification radius.

To summarize, a generic and particularly predictive setup can be described 
as follows. Assume that there are no bulk fermions or at least no 
non-$G$-universal mass splitting of bulk fermions. Assume furthermore 
that $M_S\ll M_V$, i.e., the potential stabilizing the VEV of $\Phi$ is 
relatively flat (this is generic in supersymmetry which, however, will 
be discussed in more detail below). If dangerous higher-dimension operators 
are forbidden by appropriate symmetries, the leading power correction is 
calculable and the resulting 4d gauge couplings are obtained by 
multiplying Eq.~(\ref{wf}) with the volume factor V:
\beq
\alpha_{4,i}^{-1}(M_c)=V\alpha_d^{-1}-\frac{\Gamma(2-d/2)}{6(4\pi)^{d/2-1}}
(25-d)\sum_{r_i} (VM_{V,r_i}^{d-4})T_{r_i}\,.\label{mvc}
\eeq
Here $\alpha_d^{-1}$ on the r.h. side is defined in dimensional 
regularization, which makes it independent of the subtraction scale $\mu$ 
since the coefficient of the relevant $G$-universal power-divergence 
vanishes. One may think of $\alpha_d$ as the $d$-dimensional gauge 
coupling defined at zero momentum (in complete analogy with $1/M_P$ in 
4d gravity). The $T_{r_i}$ and the relative sizes of the $M_{V,r_i}$ are 
determined by group theory (the representation of $\Phi$ and the direction 
of its VEV $v$), so that the power correction is proportional to $Vv^{d-4}$. 
The relative size of this correction is $(M_V/M)^{d-4}$. It has to be 
small enough so that even higher powers of $M_V/M$ are suppressed. 
Nevertheless, it can be significantly larger than the usual 4d GUT 
threshold corrections of relative size $\alpha_{\rm GUT}\sim 1/25$. 
Jumping somewhat ahead we would like to note that higher supersymmetry or
string theory may forbid or fix {\it all} dangerous higher-dimension 
operators and higher-loop corrections to the $d$-dimensional gauge couplings
(cf.~\cite{ddg1}), in which case one might hope to go to the region 
$M_V\sim M$ so that the relative sizes of low-energy gauge couplings are 
dominantly determined by power-law effects.

\section{Brane effects and the KK-mode approach}\label{kk}
So far, we have focussed on true bulk effects and completely neglected 
terms suppressed by powers of the bulk-size $R$. However, it is clear that 
such contributions are generically present, e.g., on the r.h. side of 
Eq.~(\ref{mvc}). One can approach this issue using $d$-dimensional 
propagators in the full, compactified geometry. However, in the present 
investigation we find it simpler to discuss these effects using an 
effective 4d framework and summing KK modes. Clearly, these two methods 
are equivalent both conceptually and quantitatively. 

To be specific, although we are prepared to neglect terms down by full powers
of $MR$ (since these terms will in general be sensitive to unknown and 
largely unconstrained brane operators), we would like to take terms into 
account that are suppressed by powers of $MR$ but enhanced by $\ln(MR)$. Such
terms are known to be important in orbifold GUTs~\cite{hn,hmr}, where they 
give rise to the calculable `differential running'~\cite{nsw} above the 
compactification scale.

For simplicity, we first consider a toy example of one extra dimension 
compactified on an $S^1$. We start with a theory with one unbroken gauge 
group $G$ and consider only the contribution of a bulk scalar with mass 
$M_{S1}\sim M_c$ in a certain representation of $G$.

Further, we compare this to a theory where the scalar mass is shifted to 
$M_{S2}\gg M_c$. The difference in the scalar contribution to the low-energy 
gauge couplings in these two models comes from the difference in 
log-contributions from the KK towers:
\bea
&&\hspace*{-3cm}\alpha_4^{-1}(M_c)_{\rm model\,2}-\alpha_4^{-1}(M_c)_{\rm 
model\,1}\nonumber
\\
&=&\frac{T_r}{24\pi}\sum_{n=-\infty}^{\infty}\left[\ln\frac{\mu^2}
{(nM_c)^2+M_{S2}^2}-\ln\frac{\mu^2}{(nM_c)^2+M_{S1}^2}\right]
\label{sel1}
\\
&\simeq&\frac{T_r}{24\pi}\left[-\ln\frac{M_{S2}^2}{M_c^2}-
2\sum_{n=1}^\infty\ln\left(1+\frac{N^2}{n^2}\right)\right]\,,\label{sel}
\eea
where $N=M_{S2}/M_c$ and $M_{S1}$ has been set to zero everywhere in 
Eq.~(\ref{sel}) except for the zero-mode contribution, where it has been 
replaced by $M_c=1/R$. This introduces only an ${\cal O}(1)$ error. The sum 
on the r.h. side of Eq.~(\ref{sel}) can be estimated as 
\beq
\sum_{n=1}^\infty\ln\left(1+\frac{N^2}{n^2}\right)\simeq \pi N-\ln N+
{\cal O}(1)\,,
\eeq
for $N\gg 1$, so that the final result reads 
\beq
\alpha_4^{-1}(M_c)_{\rm model\,2}-\alpha_4^{-1}(M_c)_{\rm model\,1}=
-\frac{T_r}{12}\,\,\frac{M_{S2}}{M_c}\,.\label{pls}
\eeq
Thus, model 2 differs from model 1 precisely by the power-like contribution
$\sim M_{S2}$, which can also be obtained from Eq.~(\ref{wf}) by setting 
$d=5$. The important point here is that one finds no additional, 
log-enhanced contribution from the momentum region above $M_c$. In other 
words, the zero-mode log merges with the KK logs to give just a pure power. 

The situation is different, however, if one compactifies on $S^1/Z_2$. In 
this case, the sum over positive and negative $n$ in Eq.~(\ref{sel1}), 
corresponding to sines and cosines, is replaced by a sum over just positive 
$n$, corresponding to cosines only (assuming positive $Z_2$ parity of the 
scalar field). The zero mode still contributes with full strength. As a 
result, the cancellation of the zero-mode log is incomplete and 
Eq.~(\ref{pls}) is replaced by 
\beq
\alpha_4^{-1}(M_c)_{\rm model\,2}-\alpha_4^{-1}(M_c)_{\rm model\,1}=
-\frac{T_r}{24}\,\,\frac{M_{S2}}{M_c}-\frac{T_r}{12\pi}\,\,
\frac{1}{2}\ln\frac{M_{S2}}{M_c}\,.
\eeq

If, on the other hand, the $Z_2$ parity of the scalar field is odd, there
is no zero mode and only the sine modes contribute to the KK sum. One then
finds
\beq
\alpha_4^{-1}(M_c)_{\rm model\,2}-\alpha_4^{-1}(M_c)_{\rm model\,1}=
-\frac{T_r}{24}\,\,\frac{M_{S2}}{M_c}+\frac{T_r}{12\pi}\,\,
\frac{1}{2}\ln\frac{M_{S2}}{M_c}\,.
\eeq

This simple calculation allows for the following intuitive interpretation:
Without branes, gauge coupling corrections are logarithmic below the 
compactification scale and purely power-like above it. Introducing 4d
boundaries (branes) leads to typical 4d effects even above $M_c$, i.e.,
logarithmic corrections. For each brane at which a 5d field is non-zero
(Neumann boundary conditions), one finds $(1/4)$ times the usual log 
from 4d running. For each brane at which a 5d field is zero
(Dirichlet boundary conditions), one finds $-(1/4)$ times this log. 
It can be easily checked that this rule extends to $S^1/(Z_2\times Z_2')$,
where a field can be zero at one brane and non-zero at the other. 

While the extension of this rule to fermions is straightforward, the case of
massive 5d vector fields requires some comments. The rule is that, if 
$A_\mu$ (where $\mu=0,...,3$) is non-zero at a boundary, one finds a scalar 
log contribution with prefactor $(1/4)(-22)$. The factor $-22$ can be 
derived from Eq.~(\ref{sel}) recalling that the zero mode (massive vector) 
has prefactor $-21$, while the KK tower (massive vectors and $A_5$-scalars) 
has prefactor $-20$. An intuitive understanding can be obtained if, guided 
by the scalar case above, one adds the $A_\mu$ contribution $(1/4)(-21)$ 
and the $A_5$ contribution $(-1/4)$ (the `$-$' arising since $A_5$ is zero 
if $A_\mu$ is non-zero). The rule extends in an obvious way to the case in 
which
$A_\mu$ is zero at a boundary (orbifold breaking of the gauge group): one 
finds a scalar log with prefactor $(-1/4)(-22)$. In deriving this, it is 
important not to forget the $A_5$ zero-mode. Furthermore, there is a
straightforward extension to the case of massless 5d vector fields, where 
the relevant prefactors of the boundary logs are $(\pm 1/4)(-23)$. 

In fact, the above set of rules represents a simple and intuitive way of 
rederiving the `differential running' in 5d orbifold GUTs above $M_c$
because it relates 4d logs directly to the boundary conditions of fields 
(without any reference to the KK mode spectrum). 

To illustrate the relevance of the above in the present context, we now
give a more complete version of Eq.~(\ref{mvc}) in 5d. We work on $S^1/Z_2$ 
with $A_\mu$ and $\Phi$ non-zero at both boundaries. The result, which now 
includes both power-law and log-enhanced terms, reads
\bea
\alpha_{4,i}^{-1}(M_c)&=&\pi R\alpha_5^{-1}+\frac{1}{24}\left[20\sum_{r_i} 
(RM_{V,r_i})T_{r_i}-\sum_{r_i'}(RM_{S,r'_i})T_{r_i'}\right]\label{ilo}
\\
&&\hspace*{-2cm}+\frac{1}{12\pi}\left[\frac{1}{2}(-22)\sum_{r_i}T_{r_i}\ln
\frac{M}{M_{V,r_i}}+\frac{1}{2}\sum_{r_i'}T_{r_i'}\ln\frac{M}{M_{S,r_i'}}
+\frac{1}{2}(-23)C_i\ln\frac{M}{M_c}\right]\,.\nonumber
\eea
Note, in particular, the appearance of contributions from the vector bosons
of the unbroken subgroup ($C_i$ is the adjoint Casimir of $H_i$) which,
although irrelevant for the power-like terms, contribute to the 
boundary-driven logarithmic running above $M_c$. Furthermore, it should
be observed that no non-universal logarithmic running occurs above the 
highest of the scales $M_{V,r_i}$ and $M_{S,r'_i}$ since
\beq
\sum_{r_i}T_{r_i}+C_i=C_A(G)=\mbox{$i$-independent}
\eeq
and
\beq
\sum_{r_i}T_{r_i}+\sum_{r'_i}T_{r'_i}=T_{\Phi {\rm -repr.}}(G)=
\mbox{$i$-independent}\,.
\eeq

\section{The supersymmetric theory}\label{sup}
Most of the above extends straightforwardly to supersymmetry. In particular, 
Eqs.~(\ref{wf}) and (\ref{ilo}) simply require the inclusion of the 
additional degrees of freedom (fermions and scalars) that are present in 
the relevant supersymmetric multiplets. However, there are also some crucial 
new points that require a separate discussion. In particular, it is 
important to understand the possible origin of the bulk VEV and the 
resulting mass spectrum, both of which are strongly constrained by SUSY. 

We first focus on 5 and 6d, where the minimal SUSY corresponds to 
$N=2$ in 4d language. This excludes all renormalizable (from the 4d 
point of view) interactions except those prescribed by gauge symmetry. In 
particular, the Higgs field $\Phi$, which would have to come from a gauged 
hypermultiplet, can not have a conventional bulk potential with cubic and 
quartic terms. Although it appears conceivable that higher-dimension 
operators, consistent with 5d SUSY, generate a suitable potential~\footnote{
A 
systematic analysis of such operators should be possible using the 
manifestly gauge-invariant formulation~\cite{heb} of 5d SUSY in terms of 
4d superfields~\cite{agw,sf}.
}
we chose the simpler option of fixing the bulk Higgs VEV by an appropriate 
boundary potential. In doing so, we follow the method for breaking 
U(1)$_\chi$ in the 6d SO(10) model of~\cite{abc}. Clearly, we have to rely 
on the existence of a D-flat direction in the bulk. (Here, by D-flatness we 
mean that no potential arises from integrating out the SU(2)-R triplet of 
auxiliary fields of the gauge multiplet. For an explicit component 
lagrangian of a gauged 5d hypermultiplet see, e.g.,~\cite{heb}.) In 
general, such a D-flat direction might not exist. This can, for example, be 
easily checked in the case of a single U(1) hypermultiplet. We now assume 
a representation or field content where a flat direction can be found. 
The non-zero VEV is stabilized only by a brane superpotential which we will 
not specify at the moment. In the bulk, the VEV will give masses to the 
whole 5d vector multiplet (in the broken directions) and to a 
whole hypermultiplet (in the directions corresponding to the would-be 
Goldstone-bosons). However, we also know that in spontaneous gauge symmetry 
breaking a single scalar degree of freedom is transferred to the vector 
field. In the case of 5d SUSY, this is only possible if the masses of the 
vector multiplet and the hypermultiplet in the broken directions are the 
same. Let us for the moment assume that these two multiplets exhaust the 
set of heavy states. 

This occurs, in particular if we choose the hypermultiplet to be in 
the adjoint representation, which makes the model $N=4$ supersymmetric. 
We can then imagine the theory to arise via dimensional reduction from a 
SYM theory in 10d and think of 
the two complex scalars of the hypermultiplet as $(A_7+iA_8)$ and $(A_9+
iA_{10})$. It is now clear that flat directions exist (e.g. $A_7=$const.)
and that the whole hypermultiplet acquires a mass (from terms $\sim 
[A_7,A_8]^2$ etc.). Furthermore, it is immediately clear from the underlying 
gauge structure that all scalar and vector masses, and hence also the 
fermionic masses, corresponding to excitations of the broken directions 
are identical. 

Thus, we have argued that, after spontaneous symmetry breaking driven by 
bulk hypermultiplets, we find the degrees of freedom of a vector 
multiplet and a hypermultiplet for every broken direction at the massive 
level. Simple counting of vector, scalar and fermionic fields according to 
Eq.~(\ref{wf}) shows that no bulk loop correction arises. This does not come 
as a surprise since we are faced with the field content corresponding to 
$N=4$ SUSY. 

However, the symmetry-breaking bulk VEV does not have to come from a 
Higgs. Instead, it is possible that, in a compact geometry, one of the 
extra-dimensional components of the vector field (e.g., $A_5$ in 5d;
$A_5$ or $A_6$ in 6d) develops a VEV. Clearly, only adjoint breaking is 
possible in this case. However, it is a well-known and difficult problem to
stabilize such a VEV. This is probably even more so if we require 
the $A_5$ or $A_6$ VEV to be large enough to generate a large 
power correction. 

A closely related and more immediately useful possibility exists in 5d. 
Consider, for example, a 5d SU(5) model. It is possible that the scalar 
partner $\Sigma$ of the gauge fields, which is present in 5d SUSY, develops 
a bulk VEV in U(1)$_Y$ direction\footnote{
We
are indebted to S. Groot Nibbelink for emphasizing this possibility in a 
very helpful conversation. 
}
(cf.~\cite{maru}).
Such a scalar VEV can arise in an $S^1/Z_2$ model where both boundaries 
break SU(5) and Fayet-Iliopoulos terms of the U(1)$_Y$ subgroup are present 
at both boundaries. As explained in~\cite{pm} (see also~\cite{agw,fi}), in 
the 5d setup this term does not break SUSY or U(1)$_Y$, but instead drives 
a non-zero bulk VEV of $\Sigma$. More generally, whenever we have a 5d 
orbifold model where the bulk gauge symmetry is broken in such a way that 
an isolated U(1) factor survives on both branes, Fayet-Iliopoulos terms 
driving a bulk VEV of $\Sigma$ can be introduced. 

In the presence of a VEV of $\Sigma$, all the fields in the 5d vector 
multiplet corresponding to the broken directions acquire a bulk mass 
$M_V$. The formula for threshold corrections relevant to this case reads
\beq
\alpha_{4,i}^{-1}(M_c)=V\alpha_d^{-1}+\frac{1}{24\pi}
12\sum_{r_i} (VM_{V,r_i})T_{r_i}\,.\label{asu}
\eeq
The prefactor 12 can be understood as the sum of 20 for a massive 5d vector 
and $-8$ for the 
spinor. The degree of freedom corresponding to $\Sigma$ is absorbed in the 
massive vector field. (As discussed above, it is also immediately clear 
that a hypermultiplet of mass $M_V$ would precisely cancel this term.) We can 
improve the correction by including volume suppressed but log-enhanced 
terms using the discussion in the previous section. For simplicity, we work 
on $S^1/Z_2$ and assume that both boundaries break the gauge group in the 
same way as the bulk VEV:
\bea
\alpha_{4,i}^{-1}(M_c)&=&\pi R\alpha_5^{-1}+\frac{1}{24}\left[12\sum_{r_i} 
(RM_{V,r_i})T_{r_i}\right]\label{suca}
\\
&&+\frac{1}{12\pi}\left[-\frac{1}{2}(-24)\sum_{r_i}T_{r_i}\ln
\frac{M}{M_{V,r_i}}+\frac{1}{2}(-24)C_i\ln\frac{M}{M_c}\right]\nonumber
\,.
\eea
Note that fermions do not contribute to the logarithmic terms since 
the two Weyl fermions contained in the 5d spinor have opposite boundary 
conditions at every brane. 

For $d=7$, the minimal vector multiplet again contains scalar adjoints that
could acquire a VEV as $\Sigma$ in the 5d case above. However, the minimal 
supersymmetry is $N=4$ in 4d language and we expect no loop corrections 
to the gauge couplings.

\section{Towards a realistic SU(5) model}\label{su5}
We now turn to a preliminary analysis of phenomenological implications of 
the power-like threshold corrections calculated above. In this section, we 
restrict ourselves to 5d SU(5) models following, in essence, the 
construction principles of the simplest orbifold GUT models~\cite{kaw,af,
hn,hmr}. Proton decay is avoided by placing fermions on branes where 
SU(5) is not a good symmetry. The light SM Higgs doublet(s) can be localized 
on the same brane, as suggested in the minimal scenario of~\cite{hmr} (which 
can be dynamically realized using bulk masses as in~\cite{hmr1}). 
Furthermore, as discussed in the previous section, we assume that the 
scalar partner $\Sigma$ of the gauge fields, which is present in 5d SUSY, 
develops a bulk VEV in U(1)$_Y$ direction. Such a scalar VEV can arise in 
an $S^1/Z_2$ model where both boundaries break SU(5) and Fayet-Iliopoulos 
terms are present. We assume that the usual problem with SU(5) models on 
$S^1/Z_2$, namely the existence of massless scalars with quantum numbers of 
the $X,Y$ gauge bosons, is solved by introducing appropriate non-local 
interactions giving these fields a mass. More precisely, the zero-modes in 
question can be understood as a chiral Wilson-line superfield~\cite{hmn}
and interactions involving this Wilson-line are naturally generated by 
integrating out massive degrees of freedom in the bulk~\cite{cgm}.

We assume a standard supersymmetric scenario in 4d, in which case the 
running between the electroweak scale and $M_c$ is the familiar MSSM 
running. The low-energy data is taken to be $\alpha_i^{-1}(m_Z)=(59.0,29.6, 
8.4)$ and the effective SUSY breaking scale is set to $m_Z$. In this case, 
the relation between couplings at $m_Z$ and $M_c$ is given by
\beq
\alpha_{4,i}^{-1}(m_Z)=\alpha_{4,i}^{-1}(M_c)+\frac{1}{12\pi}\left(-18C_i+
12T_i\right)\ln\frac{M_c}{m_Z}+\mbox{SM matter contributions}
\eeq
with $C_i=(0,2,3)$ (Casimirs of the SM gauge groups) and $T_i=(3/10,1/2,0)$ 
(SM Higgs representation). Furthermore, using the results of the previous 
sections and working on an $S^1/Z_2$, where the $Z_2$ breaks SU(5), we have
\bea
&&\hspace*{-2cm}\alpha_{4,i}^{-1}(M_c)=\pi R\alpha_5^{-1}+
\frac{1}{24}\left[12(RM_V)(5-C_i)\right]+\frac{1}{12\pi}\left[12T_i\ln
\frac{M}{M_c}\right]
\\ \nonumber\\
&&+\frac{1}{12\pi}\,\frac{1}{2}\left[24(5-C_i)\ln\frac{M}{M_V}+(-24)
C_i\ln\frac{M}{M_c}\right]\nonumber
\\ \nonumber\\
&&+\mbox{SM matter contributions}\,.\nonumber
\eea
This follows immediately from Eq.~(\ref{suca}), with the brane-localized 
Higgs contributing even above $M_c$. 

The usual fairly precise MSSM unification is formally obtained in the 
limit $M=M_c=M_V=M_{\rm GUT}$. We can now try to lower $M_c$ and see 
whether we can maintain gauge unification at the cost of the power law 
term $\sim M_V/M_c$. This is not hopeless because the coefficients $-C_i$ 
coming with this term represent the main part of the usual MSSM 
running coefficients. We focus on differences of 4d inverse gauge 
couplings, $\alpha_{ij}\equiv \alpha_i^{-1}-\alpha_j^{-1}$. The crucial 
gauge unification constraint can be characterized by 
\beq
\frac{\alpha_{12}(m_Z)}{\alpha_{23}(m_Z)}=\frac{59.0-29.6}{29.6-8.4}=1.39
\,\,.\label{a12}
\eeq
This has to be compared with the result obtained from combining the above
running and threshold formulae:
\bea
&&\hspace*{-2cm}\alpha_{ij}(m_Z)=\frac{1}{12\pi}\left(-18C_{ij}+12T_{ij}
\right)\ln\frac{M_c}{m_Z}\label{aij}
\\
&&+\frac{1}{24}\left[-12C_{ij}\right]\frac{M_V}{M_c}+\frac{1}{12\pi}\left[
12T_{ij}-12C_{ij}\right]\ln\frac{M}{M_c}-\frac{1}{12\pi}12C_{ij}\ln\frac{M}
{M_V}\,,\nonumber
\eea
where $C_{ij}=C_i-C_j$ and $T_{ij}=T_i-T_j$. 

The maximal value of $M$ suggested by NDA~\cite{nda} (cf.~\cite{hmr}) can 
be characterized by $M/M_c\sim 10^3$.
The validity range of our calculation is $M_c\ll M_V\ll M$. It is amusing 
to observe that, if we set $M_V\simeq \sqrt{M_cM}$ to realize this 
situation, the logarithmic terms from the energy range above $M_c$ 
mimick precisely the MSSM contribution to coupling ratios. Thus, even 
if $M_V$ does not have this precise value, the log terms will not affect 
MSSM-type unification significantly and, given the preliminary character
of the present investigation, we now focuss on the power term.
From Eq.~(\ref{aij}) one can read off that just the logarithmic MSSM 
contribution would give $\alpha_{12}/\alpha_{23}=1.4$ while just the 
power-like term would give $\alpha_{12}/\alpha_{23}=2$ (cf.~Eq.~(\ref{a12})).
Thus, to maintain the above field content while lowering the unification 
scale significantly, one has to sacrifice precision. 
One can expect to find $\alpha_{12}/\alpha_{23}=1.5$ at $m_Z$ if about 
$1/6$ of the log-running is traded for the power correction. This lowers 
the unification
scale $M$ to about $10^{14}$ GeV thus allowing, e.g., for a see-saw mechanism 
based directly on the GUT scale (without the usual mismatch by a factor 
${\cal O}(10)$). One could also consider the possibility that there are no
right-handed neutrinos and light neutrino masses are based directly on the 
appropriate higher-dimension operator suppressed by the new GUT scale. 
However, the price to pay is the extra ${\cal O}(1)$ threshold corrections 
to $\alpha_i^{-1}$ that are needed for consistency with the low-energy data. 
Although such corrections are not unnatural, given that a significant 
log-running continues all the way up to UV-scale $M$, they are certainly 
larger than what would be needed in the 4d MSSM. 

Next, we want to consider the possibility that power-like threshold 
corrections beyond those driven by the 5d gauge multiplet arise. This would 
not be possible if the bulk breaking was realized by a bulk Higgs field 
since 5d SUSY forbids the necessary coupling of this Higgs with other 
hypermultiplets. However, since we consider gauge breaking by the scalar 
adjoint, this possibility exists. If we add a bulk hypermultiplet, say in 
the 5 of SU(5), then the doublet and triplet part of it acquire different 
bulk masses due to the 
coupling to $\Sigma$. The ratio of $M_V$ and these two masses $M_d$ and 
$M_t$ is prescribed by elementary group theory:
\beq
M_V:M_d:M_t\sim 5:3:2\,.
\eeq
The power-like threshold corrections arising in this situation read
\beq
\Delta\alpha_{4,i}^{-1}(M_c)=\frac{1}{24}\,\frac{M_V}{M_c}\left[12(5-C_i)
-12(\frac{3}{5}T_i+\frac{2}{5}T'_i)\right]\,,\label{del1}
\eeq
where $T'_i=(2/10,0,1/2)$ characterize the Higgs triplet representation. 
On the basis of just this power-law contribution one would have 
$\alpha_{12}/\alpha_{23}\simeq 2.27$, i.e., a situation worse than 
without the bulk 5. 

However, this effect can be turned to its opposite by also 
introducing an SU(5)-invariant bulk mass $M_f$ for the 5-hypermultiplet. 
In the presence of such a mass, quantified by $\xi=M_f/M_V$, with 
$\xi\simeq 0.4$ and with the sign chosen such that it almost compensates 
the $\Sigma$-driven doublet mass, Eq.~(\ref{del1}) is transformed into
\beq
\Delta\alpha_{4,i}^{-1}(M_c)=\frac{1}{24}\,\frac{M_V}{M_c}\left[12(5-C_i)
-12\left(\left|\frac{3}{5}-\xi\right|T_i+\left|\frac{2}{5}+\xi\right|T'_i
\right)\right]\,,\label{del2}
\eeq
leading to $\alpha_{12}/\alpha_{23}\simeq 1.44$ just from the power-like 
correction. Now power-like threshold corrections can replace a significant 
part of the MSSM log-running without loss of precision of unification, but 
at the cost of tuning $M_f$. (This tuning can, of course, also be used to 
achieve perfect unification, including even the brane-driven log-running 
above $M_c$.) Dangerous additional terms can come from higher-dimension 
operators. In particular, an operator\footnote{
In contrast to the case where the VEV comes from a hypermultiplet, the 
$\Sigma$-VEV can also couple linearly, $\sim F^2\Sigma$, as in the 
super-Chern-Simons term discussed in~\cite{agw}. Here we assume that this 
term is either forbidden or small. However, even if this term is
required (e.g. to cancel anomalies at the boundary), its presence does not
destroy the predictivity of the scenario because its contribution to 
low-energy gauge coupling differences is prescribed by simple group theory.
}
$\sim F^2\Sigma^2$ can contribute to 
$\alpha_4^{-1}$ at the level $M_V^2/(M_cM)$. If we require this term not to
be larger than ${\cal O}(1)$ and take $M\sim 10^3M_c$, we find the 
constraint $M_V\lsim 30M_c$. (More optimistically, one could assume that 
this term is forbidden or at least uniquely specified in its structure by
$N=2$ SUSY.) From Eq.~(\ref{del2}) we can now read off that
about half of the low-energy value of, say, $\alpha_{12}$ can be due to
power-like term, so that $M$ and $M_c$ can be lowered to $\sim 10^9$ GeV and 
$\sim 10^6$ GeV respectively. Given that our very crude estimates have 
produced this quite impressive result, a more detailed numerical study,
including two-loop running and considering appropriate NDA factors, appears 
to be warranted. 

Going further, one might even consider that an exact or almost exact 
cancellation, based on some yet unknown symmetry reason, leads to vanishing 
bulk doublet mass $(\xi=3/5)$. Equation~(\ref{del2}) then implies 
$\alpha_{12}/\alpha_{23}\simeq 1.20$, which is close enough to the desired 
value $1.39$ to have considerable power-law effects without a significant 
loss of unification precision. In fact, it is this specific scenario that
comes closest to original proposal of~\cite{ddg}. The price for
this is the re-emergence of the familiar SU(5) problem of tuning the 
doublet mass to zero. Furthermore, one gets only a single Higgs-doublet at 
the zero mode level from the one bulk hypermultiplet. Two bulk 
hypermultiplets would, unfortunately, lead to a much stronger deviation 
from the desired low-energy coupling ratio. 

Let us, however, note that both MSSM Higgs doublets can come from the same 
hypermultiplet as the two boundary-localized massive modes with 
exponentially suppressed 4d mass. This requires the bulk doublet mass to 
be sufficiently large rather than small. With the above favoured value
$\xi\simeq 0.4$ and the maximal allowed vector mass $M_V\simeq 30M_c$, we 
find a bulk doublet mass $M_d\simeq 6 M_c$. The effective 4d Dirac mass 
linking the two boundary modes is $m_4\simeq 2M_d\exp(-M_d\pi R)\sim
10^{-8}M_d$ (see, e.g.,~\cite{pm,hmr1}), which may be acceptable in 
settings with very low unification scale.

It certainly would be interesting to extend the above preliminary analysis 
to various other proposals involving gauge unification in extra dimensions 
where power-law effects can be important (see, for example, the discussion 
in~\cite{ll} and~\cite{pst}. However, this is beyond the scope of the 
present paper. 

Before closing, we now turn to the possible role of power-like threshold 
corrections in non-supersymmetric 5d SU(5) models. For this purpose, we 
choose to accept an ad-hoc fine-tuning solution of the well-known problem 
of quadratically divergent Higgs mass corrections and focus exclusively on 
the precision of gauge coupling unification. 

We consider an $S^1/(Z_2\times Z_2')$ model with SM fermions and Higgs 
doublet on the SU(5)-breaking brane. The bulk Higgs field $\Phi$ is in the 
adjoint of SU(5) and develops a VEV in U(1)$_Y$ direction. The 4d running 
below $M_c$ gives $\alpha_{12}/\alpha_{23}=1.90$, in 
significant disagreement with data. To correct this, we consider power-like 
threshold corrections from the bulk, introducing a set of fundamental 
fermions of SU(5) coupled to the adjoint Higgs by a standard Yukawa 
coupling $\sim\bar{\psi}\Phi\psi$. Since these fermions can also have an 
SU(5) symmetric bulk mass, we can treat the resulting doublet and triplet 
masses $M_{\psi,t}$ and $M_{\psi,d}$ as essentially independent parameters. 
Assuming that $\Phi$ has no or only a very small bulk mass, the non-SUSY 
analogue of Eq.~(\ref{aij}) reads
\bea
&&\hspace*{-1.3cm}\alpha_{ij}(m_Z)=\frac{1}{12\pi}\left(-22C_{ij}+2T_{ij}
\right)\ln\frac{M_c}{m_Z}
\\
&&+\frac{1}{48}\left[-20C_{ij}\frac{M_V}{M_c}-8T_{ij}\frac{M_{\psi,d}}{M_c}
-8T'_{ij}\frac{M_{\psi,t}}{M_c}\right]+\frac{1}{12\pi}\left[2T_{ij}+
\frac{1}{2}(-22)C_{ij}\right]\ln\frac{M}{M_c}\,.\nonumber
\eea
For simplicity, let us assume $M_{\psi,d}\ll M_{\psi,t}$ so that the 
power-like contribution from the doublet can be neglected. Further, we 
lower the compactification scale as far as possible according to the naive 
estimate based on the higher-dimension operator discussed above, 
$M_c\simeq 10^6$ GeV with $M_V\simeq 20M_c$ and $M\simeq 10^3M_c$. One 
now finds that the moderate value $M_{\psi,t}\simeq 3M_V$ gives 
$\alpha_{12}(m_Z)\simeq 29.4$ and $\alpha_{23}(m_Z)\simeq 21.2$, in 
reasonable agreement with the data. Although, given the ad-hoc choice of
several parameters, this certainly does not challenge the numerical 
superiority of the minimal SUSY framework, it is nevertheless interesting 
to see how easily a non-SUSY SU(5) unification can be achieved with the 
help of large power-like thresholds.

Given that the above exploratory study has shown the possibility of 
very low compactification scales $\sim 10^6$ GeV, it is tempting to 
speculate that further work and a better understanding of the 
UV theory will reveal viable scenarios with TeV scale precision 
unification.

\section{Conclusions}\label{con}

In this paper, we have analyzed the role of loop corrections to gauge 
coupling constants in grand unified theories with more than 4 dimensions. 
Since such theories are non-renormalizable, these corrections are, in 
general, UV-dominated. However, if the higher-dimensional theory respects 
a certain large gauge symmetry, which is softly broken in the perturbative 
domain, the differences of the gauge couplings of the surviving subgroups 
can be calculable and independent of the UV completion. This is obvious in 
cases where the relevant counterterm is forbidden by the symmetries of 
the fundamental theory and can also be checked by an explicit analysis of 
the relevant loop integrals.

More specifically, in 5d gauge theories softly broken at a scale $M_B$, 
differences of inverse low-energy gauge couplings receive finite corrections 
$\sim M_B/M$ (where $M$ is the UV scale set by the dimensionful 5d gauge 
coupling). If the relevant Higgs field can not appear linearly in the 
lagrangian for symmetry reasons, higher-dimension operators can not compete 
with these loop corrections. In 6d, the corresponding correction is 
$\sim M_B^2/M^2\ln \Lambda$, where the $\Lambda$ is the UV cutoff. This 
logarithmic divergence demands the existence of an appropriate 
higher-dimension 
operator (which has to be quadratic in the Higgs field). Nevertheless, at 
least at the leading logarithmic level, calculability is not lost. In seven 
and more dimensions, non-universal (with respect to the low-energy gauge 
groups) power-divergent corrections are expected to arise in 
non-supersymmetric models. Thus, quantitative statements depend on 
a more detailed knowledge of the UV structure of the theory. However, 
group theoretical constraints on the relevant higher-dimension operators 
may be sufficient to characterize the way in which these terms can 
contribute to low-energy gauge coupling differences. 

A crucial feature of the supersymmetric theory is the strong restriction 
that the higher supersymmetry places on the bulk Higgs potential. A 
non-trivial bulk Higgs VEV can, however, be enforced by an appropriate 
brane superpotential. For adjoint Higgs breaking, power-like threshold 
corrections do not arise because, even in 5 and 6d, the massive field 
content is that of $N=4$ super-Yang-Mills theory. 
However, if the breaking is due to a bulk VEV of the scalar adjoint from 
the vector multiplet, potentially large power corrections arise. 
Further important issues, which have 
not been discussed in the present paper (see, however,~\cite{ddg1}) are the 
$N=2$ SUSY restrictions on higher-loop terms and the possibility of a 
non-trivial UV fixed point (see~\cite{fp} for the UV fixed point structure 
in non-SUSY gauge theories in higher dimensions). In particular, it is 
possible that SUSY constraints on higher-dimension operators and 
control of higher-loop corrections will allow a significant extension of 
the calculability range $(M_V\ll M)$ assumed in this paper. Such a more 
detailed knowledge can, of course, also arise from a successful embedding 
of the GUT scenario in string or M theory (see~\cite{wit} for a very recent 
investigation), where threshold effects are known to be calculable, and 
lead to large and highly-predictive power-law effects. 

In the phenomenological part of the present paper, we have focussed on 
supersymmetric SU(5) models in 5d. In the simplest setting, where only the 
bulk gauge multiplet contributes power-like thresholds, the deviation 
from MSSM running is considerable and the compactification scale $M_c$ can 
only be lowered moderately (say, down to a phenomenologically favoured 
neutrino see-saw scale). Since the symmetry breaking bulk VEV does not come 
from a Higgs hypermultiplet but from the scalar adjoint of the gauge 
multiplet, gauged bulk matter can contribute to the power-law effects. 
In this case, we find models that are numerically very close to the 
original proposal of~\cite{ddg}. This can be understood as follows: 
in the low-energy approach, one sums KK modes of light fields; in our UV
based approach, one finds loop corrections proportional to the mass of
heavy 5d fields. However, in an SU(5) symmetric bulk these two sets of 
fields combine to full SU(5) multiplets and the relevant group theoretical 
coefficients complement each other in such a way that the effect on 
low-energy gauge coupling differences is consistent in both approaches. 
Finally, we have found that it is possible to achieve non-SUSY unification 
using extra matter content above $M_c$ and to lower the unification 
scale significantly both in the SUSY and non-SUSY case.

To summarize, we believe that the presented calculations strongly encourage 
the further quantitative study of power-like threshold corrections in 
various phenomenologically relevant models. However, it is crucial, both at 
a qualitative and at a numerical level, to start from a bulk theory with 
manifest unified gauge symmetry and to specify the details of the soft 
higher-dimensional breaking mechanism as well as the field content at the 
high scale. 

\vspace*{.3cm}
While this paper was being typed, Ref.~\cite{ads1} appeared, which considers 
warped 5d models and has some overlap with our results.

\noindent {\bf Note added}: 
In~\cite{seib,intr}, loop corrections to 5d SYM theories were calculated
within the prepotential formalism of the corresponding 4d $N=2$ theory.
Since the prepotential is only corrected at one loop and higher-dimension 
terms are forbidden by 5d gauge invariance, a quantum-exact prepotential 
could be obtained. From this quantity, the power-like threshold 
corrections considered in this paper can be extracted. This represents an 
alternative to our component analysis on the basis of early non-SUSY GUT 
threshold calculations. More importantly, the quantum-exactness of the 
one-loop prepotential implies that our analysis is, in fact, complete and 
can therefore be taken to the strong coupling region. Thus, TeV-scale 
precision unification becomes a realistic possibility. We are indebted to 
Erich Poppitz for drawing our attention to Refs.~\cite{seib,intr} some time 
after this paper was published.

\noindent
{\bf Acknowledgements}: We are grateful to Wilfried Buchm\"uller, 
Stefan Groot Nibbelink, John March-Russell, Riccardo Rattazzi, Dominik 
St\"ockinger and Taizan Watari for helpful discussions.

\end{document}